\documentclass[useAMS,usenatbib]{mn2e}
\usepackage{amsmath}
\usepackage{amssymb}
\usepackage{graphics}
\usepackage{graphicx}
\usepackage{epsfig} 
\usepackage{hyperref}
\def\be{\begin{equation}}
\def\ee{\end{equation}}
\def\ba{\begin{eqnarray}}
\def\ea{\end{eqnarray}}

\usepackage{color}
\definecolor{red}{rgb}{1,0.0,0.0}

\definecolor{blue}{rgb}{0.1,0.3,0.9}

\usepackage[normalem]{ulem}
\definecolor{darkgreen}{rgb}{0.0,0.5,0.0}

\newcommand{\documentname}{paper~}

\newcommand{\beq}{\begin{eqnarray}} 
\newcommand{\eeq}{\end{eqnarray}}

\newcommand{\apj}{ApJ} 
\newcommand{\apjs}{ApJS} 
\newcommand{\apjl}{ApJL} 
\newcommand{\aj}{AJ} 
\newcommand{\mnras}{MNRAS}

\newcommand{\nat}{Nature}

\newcommand{\ly}{{\ifmmode{{\rm Ly}\alpha}\else{Ly$\alpha$}\fi}}
\newcommand{\hMpc}{{\ifmmode{h^{-1}{\rm Mpc}}\else{$h^{-1}$Mpc }\fi}} 
\newcommand{\hGpc}{{\ifmmode{h^{-1}{\rm Gpc}}\else{$h^{-1}$Gpc }\fi}} 
\newcommand{\hmpc}{{\ifmmode{h^{-1}{\rm Mpc}}\else{$h^{-1}$Mpc }\fi}} 
\newcommand{\hkpc}{{\ifmmode{h^{-1}{\rm kpc}}\else{$h^{-1}$kpc }\fi}}
\newcommand{\hMsun}{{\ifmmode{h^{-1}{\rm
        {M_{\odot}}}}\else{$h^{-1}{\rm{M_{\odot}}}$~}\fi}}  
\newcommand{\hmsun}{{\ifmmode{h^{-1}{\rm
        {M_{\odot}}}}\else{$h^{-1}{\rm{M_{\odot}}}$}\fi}}  
\newcommand{\Msun}{{\ifmmode{{\rm {M_{\odot}}}}\else{${\rm{M_{\odot}}}$}\fi}} 
\newcommand{\msun}{{\ifmmode{{\rm {M_{\odot}}}}\else{${\rm{M_{\odot}}}$}\fi}}

\newcommand{\rand}{{\ifmmode{{\mathcal{R}}}\else{${\mathcal{R}}$ }\fi}} 

\newcommand{\muavg}{\vert\langle\cos\theta\rangle\vert}
\def\lsim{~\rlap{$<$}{\lower 1.0ex\hbox{$\sim$}}}

\def\gsim{~\rlap{$>$}{\lower 1.0ex\hbox{$\sim$}}}

\begin{document}

\title[Halo alignments with the cosmic web]{Cosmic web alignments with
  the shape, angular momentum and peculiar velocities of dark matter
  haloes}  
\author[J.E. Forero-Romero et al.]{
\parbox[t]{\textwidth}{\raggedright
  Jaime E. Forero-Romero$^{1}$,
  Sergio Contreras$^{2,3}$,
  Nelson Padilla$^{2,3,4}$
}
\vspace*{6pt}\\
$^{1}$Departamento de F\'{i}sica, Universidad de los Andes, Cra. 1
No. 18A-10, Edificio Ip, Bogot\'a, Colombia\\
$^{2}$Instituto de Astrofísica, Pontificia Universidad Cat\'olica,
Av. Vicu\~na Mackenna 4860, Santiago, Chile\\
$^{3}$Centro de Astro-Ingenier\'\i a, Pontificia Universidad Cat\'olica,
Av. Vicu\~na Mackenna 4860, Santiago, Chile\\
$^{4}$Max Planck Institut f\"ur Astrophysik, Postfach 1317, D-85741, Garching, Germany\\
}
\maketitle

\begin{abstract}
We study the alignment of dark matter haloes with the cosmic web
characterized by the tidal and velocity shear fields. We focus on the
alignment of their shape, angular momentum and peculiar velocities. We  
use a cosmological N-body simulation that allows to study dark matter
haloes spanning almost five orders of magnitude in mass
($10^{9}$-$10^{14}$) \hMsun and spatial scales of
$(0.5$-$1.0)$\hMpc to define the cosmic web. The strongest alignment
is measured for halo shape along the smallest tidal eigenvector,
e.g. along filaments and walls, with a signal that gets stronger as
the halo mass increases. In the case of the velocity shear field only
massive haloes $>10^{12}$\hMsun tend to have their shapes aligned along
the largest tidal eigenvector; that is, perpendicular to filaments and
walls. For the angular momentum we find alignment signals only for
haloes more massive than $10^{12}$\hMsun both in the tidal and velocity
shear fields where the preferences is to be parallel to the
middle eigenvector; perpendicular to filaments and parallel to
walls. Finally, the peculiar velocities show a strong alignment along
the smallest tidal eigenvector for all halo masses; haloes move along filaments
and walls. The same alignment is present with the velocity shear,
albeit weaker and only for haloes less massive than $10^{12}\hMsun$. Our
results show that the two different algorithms used to
define the cosmic web describe different physical aspects of
non-linear collapse and should be used in a complementary way to
understand the cosmic web influence on galaxy evolution.    
\end{abstract}

\section{Introduction}
\label{sec:introduction}

There is a long observational tradition studying galactic properties
as a function of their large scale environment
\citep[e.g.][]{Oemler1974,Dressler1980,Pimbblet2002,Gomez2003,Kauffmann2004,Abbas2006,Baldry2006,Park2007,OMill2008,Gonzalez2009,Padilla2010,Wilman2010,Muldrew2012}. In these situations the environment definition is usually based on
quantities easily accessible to observations such as local number
density or nearest neighbour measurements. With
the advent of large galaxy surveys and cosmological N-body simulations
the visibility of the cosmic web and its physical origin became
clear. As a consequence, in order to capture its filamentarity, the
environment definition started to be more complex, involving shear and gradient properties of the galaxy density field or the reconstructed/simulated dark
matter density/velocity field
\citep[e.g.][]{Lee2005,Basilakos2006,AragonCalvo2007,Hahn2007,Sousbie2008,Zhang2009,Tweb,MunozCuartas2011,Vweb,Trowland2013,Tempel2014},
including recent developments that take into account the rotational
part of the velocity field \citep[e.g.][]{Wang2013,LibeskindVorticity}.

In parallel to the observational advances, numerical simulations
successfully reproduced the web-like structure of the galaxy
distribution in models based on gravitational instability in a Dark
Matter (DM) dominated universe
\citep[e.g.][]{Bond1996,Colberg2005}. Simulations now allow to follow
the time evolution into the deep non-linear regime of virialized
structures (DM haloes) which in turn should host observable
galaxies to study their evolution in the cosmic web. The discovery in
simulations of gas filaments that feed galaxies is another theoretical
hint that strengthened the expected connection between galactic
properties and their place in the web \citep{Ocvirk2008,Dekel2009}.  

In the last decade new algorithms have implemented cosmic web
classifications that go beyond the local density by defining a
location to be a peak, sheet, filament or void depending on the
symmetry properties of the local dark matter distribution. With the
aid of simulations it has been established that haloes of a given mass
form earlier in denser regions; concentration, angular momentum and
shape can also be aligned with these web
elements. \citep[e.g][]{AragonCalvo2007,Hahn2007,Zhang2009,Gonzalez2010,Noh2011,Codis2012,Libeskind2013,Trowland2013}.     

The observational studies that try to measure angular momentum correlations
use the galaxy shape as a proxy  \citep{Lee2002,Lee2007,Jones2010,Paz2008}. In
this respect it is useful to have a theoretical baseline for the
correlations of angular momentum and shape with the cosmic web.  There
is large tradition of alignment measurements of shape and angular momentum
\citep[e.g.][]{Hahn2007,AragonCalvo2007,Zhang2009,Paz2010,Codis2012,Trowland2013,Libeskind2013,AragonCalvo2014}. The
main result of these previous studies is that shape alignment is a
robust measurement regardless of the methods and simulations. On the
other hand, the results for the angular momentum differ in the degree
of the alignment.  

In the same spirit of describing the place of galaxies within the
cosmic web, there has been a revival of surveys that measure the
cosmic flow patterns in the Local Universe
\citep{Nusser2011,Tully2013}. Assuming the linearity between the
divergence of the cosmic flow velocity field and the local matter
overdensity (valid in the linear regime) one could construct accurate
maps of the matter density from peculiar velocities
\citep{Courtois2012}. From this perspective it is interesting to look
at the expected alignment of the peculiar velocities.

In this \documentname we review most of the studies about shape and
angular momentum alignment and offer our own study with complementary
numerical techniques and simulations.  We also present for the first
time in the literature new results for the alignment of peculiar
velocities with the large scale structure.

The structure of this \documentname is the following. In \S
\ref{sec:theory} we present the theoretical antecedents for the
alignment studies we present in this paper. In \S
\ref{sec:nbody} we present the N-body
cosmological simulation and halo catalogs.  Next, we describe in
\S\ref{sec:algorithms} the two web-finding algorithms we use and in
\S\ref{sec:experiments} the setup for out numerical experiments. In
\S\ref{sec:results} we present our main results about the alignment
of shape, angular momentum and peculiar velocities with respect to the cosmic
web. In \S\ref{sec:conclusions} we present our conclusions.

\section{Theoretical Considerations: Notation and Precedents}
\label{sec:theory}

Out of the three alignments that we study in this paper -shape,
angular momentum and peculiar velocity- only the the first two have
received wide attention in the literature, being angular momentum the
most popular with twice the number of studies for shape alignment.

In this paper we focus our attention on results published during the last
decade that have made use of large N-body dark matter only
cosmological simulations. There are many works that have addressed
this problem using observational data from large surveys such as the
Sloan Digital Sky Survey (SDSS), however we choose to narrow our
discussion to simulation based studies which are
directly comparable to the one we present here.

Alignments are often measured from the distribution of the
$\mu=\vert\cos\theta\vert$ where $\theta$ is the angle between the two axes of
interest. This is often directly measured as the absolute value of
the dot product between the two unit vectors along the directions
being tested. For instance, in the case of angular momentum one would compute
$\mu=\vert\hat{j}\cdot\hat{n}\vert$. In the case of shape alignments the major
axis is the chosen direction to compare against the cosmic web.

For an isotropic distribution of the vector around the direction defined by
$\hat{n}$ the $\mu$ distribution, ranging between $0$ and $1$, should
be flat and its mean value should be
$\langle|mu|\rangle=0.5$. If a distribution is biased towards $1$
($\langle\vert\mu\vert\rangle>0.5$) we call this an statistical
alignment along $\hat{n}$, while in the case of a bias towards $0$
($\langle\vert\mu\vert\rangle<0.5$) we talk about an anti-alignment,
meaning a perpendicular orientation with respect to the $\hat{n}$
direction.

\cite{Trowland2013} presented a parametrization for the $\mu$
distribution in the case of angular momentum alignment based on
theoretical considerations by \cite{Lee2005} (equation \ref{eq:distro} in
Appendix A). Under this parametrization a unique
correspondence was found between the full shape of the $\vert\mu\vert$
distribution and its average. In this paper we follow the lines of their work but
only present the results for the average $\langle\vert\mu\vert\rangle$.

Tables 1 and 2 summarize recent results found in the literature for
shape and angular momentum alignment. Appendix A includes a detailed
description of the definitions, algorithms and simulations used in
each one of these studies. In these tables the first column describes
the reference, the second column summarizes the web finding method
with a single name, the third associates a spatial scale to the
web finding methods, in most cases it corresponds to the grid size or
smoothing scale used to interpolate the underlying matter density or
velocity field; The fourth column indicates along which web element
(filament or wall) the alignment was measured; the fifth column
indicates the strength of the alignment/anti-alignment, $++$/$--$
indicate a strong alignment/anti-alignment while $+$/$-$ indicate a
weaker signal; the last column indicates whether the described signal
is present within a defined range of halo mass.

These results can be summarized in three important points:
\begin{itemize}
\item The halo mass of $1-5\times 10^{12}$\hMsun is a threshold mass between
behaviours of no-alignment, alignment or anti-alignment.
\item Halo shape provides a strong alignment signal along filaments
  and sheets, more so for massive haloes.
\item Halo angular momentum tends to be oriented perpendicular to filaments and
parallel to sheets, but it is weaker than shape alignment.
\end{itemize}

A novel aspect of our study is the use of a single computational
volume of a high resolution simulation to study the
alignments. Equally important, is our focus to quantify to what extent
these results depend on the method used to define the cosmic web and
the numerical choices to implement the algorithms.  

\begin{table*}
\begin{tabular}{cccccc}\hline\hline
Author & Web method & Spatial scale & Along &
Alignment & Mass dependence\\
 & & ($\hMpc$)& & & \\\hline

{\bf Forero-Romero et al. (2014)} & T-Web & $0.5-1$ &
$\hat{e}_3$ (filament) &$++$ & $>10^{12}$\hMsun\\
{\bf (This Work)}&   & &
$\hat{e}_3$ (filament) & $+$ & $<10^{12}$\hMsun\\

&   & &
$\hat{e}_1$ (wall) & $++$ & $>10^{12}$\hMsun\\

&   & &
$\hat{e}_1$ (wall) & $+$ & $<10^{12}$\hMsun\\\hline

{\bf Forero-Romero et al. (2014)} & Vp-Web & $0.5-1$ &
$\hat{e}_3$ (filament) &$--$ & $>10^{12}$\hMsun\\
{\bf (This Work)}&   & &
$\hat{e}_3$ (filament) & none & $<10^{12}$\hMsun\\
&   & &
$\hat{e}_1$ (wall) & $--$ & $>10^{12}$\hMsun\\

&   & &
$\hat{e}_1$ (wall) & none & $<10^{12}$\hMsun\\\hline

\cite{Libeskind2013} & V-Web & $1$ &
filament &$++$ & $>10^{12}$\hMsun\\
&   & &
filament &$+$ & $<10^{12}$\hMsun\\
&   & &
wall & $++$ & all masses\\\hline

\cite{Zhang2009}  & Hessian density field &  $2.1$ &
filament & $++$ & $>10^{12}\hMsun$\\

& &  &
filament & $+$ & $<10^{12}\hMsun$\\\hline

\cite{AragonCalvo2007} & Hessian density field & - &
wall & $++$ & $>10^{12}$\hMsun\\

& & - &
wall & $+$ & $<10^{12}$\hMsun\\

& & - &
filament& $++$ & $>10^{12}$\hMsun\\

& & - &
filament& $+$ & $<10^{12}$\hMsun\\\hline \hline

\end{tabular}\\
\caption{Shape alignment with the cosmic web, 
$\hat{e}_1$ ($\hat{e}_3$)
is the major (minor) eigenvector of
the corresponding tensor. 
Summary of theoretical
  results provided by methods similar to ours.}
\end{table*}

\begin{table*}
\begin{tabular}{cccccc}\hline\hline
Author & Web Method & Spatial Scale& Along &
Alignment & Mass dependence\\
 & & ($\hMpc$)& & & \\\hline

{\bf Forero-Romero et al. (2014)} & T-Web & $0.5-1$ &
$\hat{e}_3$  (filament) & $-$ & $>10^{12}$\hMsun\\

{\bf (This Work)}&   & &
$\hat{e}_3$  (filament) & none & $<10^{12}$\hMsun\\

&   & &
$\hat{e}_1$ (wall) & none & $>10^{12}$\hMsun\\

&   & &
$\hat{e}_1$ (wall) & none & $<10^{12}$\hMsun\\\hline

{\bf Forero-Romero et al. (2014)} & Vp-Web & $0.5-1$ &
$\hat{e}_3$  (filament) & none & $>10^{12}$\hMsun\\

{\bf (This Work)}&   & &
$\hat{e}_3$ (filament) & none & $<10^{12}$\hMsun\\

&   & &
$\hat{e}_1$ (wall) & + & $>10^{12}$\hMsun\\

&   & &
$\hat{e}_1$ (wall) & none & $<10^{12}$\hMsun\\\hline

\cite{Libeskind2013} & V-Web & $1$ &
filament &$-$ & $>10^{12}$\hMsun\\

&   & &
filament &$+$ & $<10^{12}$\hMsun\\

&   & &
wall & $++$ & all masses\\\hline

\cite{Trowland2013} & Hessian density & $2-5$ &
filament & $-$ & $> 5\times 10^{12}$\hMsun\\
&   & &
filament & $+$ & $< 5\times 10^{12}$\hMsun\\\hline

\cite{Codis2012} & Morse Theory \& T-Web & $1-5$ &
filament & $--$ & $>10^{12.5}$\hMsun \\

&   & &
filament & $++$ & $<10^{12.5}$\hMsun \\

& & &
wall & $++$ & all masses\\\hline

\cite{Zhang2009}  & Hessian density &  $2.1$ &
filament & $++$ & if anticorrelated with shape\\

& &  &
filament & $--$ & if correlated with shape\\\hline

\cite{AragonCalvo2007} & Hessian density & - &
wall & $++$ & $>10^{12}$\hMsun\\

& & - &
wall & $+$ & $<10^{12}$\hMsun\\

& & - &
filament& $-$ & $>10^{12}$\hMsun\\

& & - &
filament& $+$ & $<10^{12}$\hMsun\\\hline

\cite{Hahn2007} & Tidal Web & $2.1$ & filament & $-$& none\\

& & &
wall & $++$ & $>10^{12}$\hMsun\\
& &    &
wall& $+$ & $<10^{12}$\hMsun\\\hline \hline

\end{tabular}
\caption{Angular momentum alignment with the cosmic web, 
$\hat{e}_1$ ($\hat{e}_3$)
is the major (minor) eigenvector of
the corresponding tensor. 
Summary of theoretical
  results provided by methods similar to ours.}

\end{table*}

\section{N-body simulation and halo catalogue}
\label{sec:nbody}

In this \documentname we use the Bolshoi simulation that follows the
non-linear evolution of a dark matter density field on cosmological
scales. The volume is a cubic box with 250\hMpc on a side, the matter
density field is sampled with $2048^3$ particles. The 
cosmological parameters in the simulation correspond to the results
inferred from WMAP5 data \citep{2009ApJS..180..306D}, which are also consistent with the more
recent results of WMAP9 \citep{2013ApJS..208...19H}. These parameters are $\Omega_m=0.27$,
$\Omega_{\Lambda} =0.73$, $\sigma_8=0.82$, $n_s=0.95$ and $h=0.70$ for the
matter density, cosmological constant, normalization of the power
spectrum, the slope in the spectrum of the primordial matter
fluctuation and the dimensionless Hubble constant. With these
conditions the mass of each dark matter particle in the simulation
corresponds to $m_p=1.4\times 10^{8}$\hMsun. A more detailed
description of the simulation can be found in
\citep{2011ApJ...740..102K}.

In this paper we use groups identified with a Friends-Of-Friends (FOF) halo
finder using a linking length of $b=0.17$ times the mean interparticle
separation. This choice translates into haloes with a density of $570$
times the mean density at $z=0$. The measurements for the shape,
angular momentum and peculiar velocity are done using the set of
particles in each dark matter halo. The definition we use in this
\documentname for the shape comes from the diagonalization of the
reduced inertia tensor.

\begin{equation}
{\mathcal T}_{lm} = \sum_{i}\frac{x_{i,l}x_{i,m}}{R_i^2},
\end{equation}
where $i$ is the particle index in the halo and $l,m$ run over the
three spatial indexes  and $R_i^2 = x_{i,1}^2 + x_{i,2}^2 +
x_{i,3}^2$, where the positions are measured with respect to the
centre of mass.

The angular momentum is calculated as
\begin{equation}
\vec{J} = \sum_{i}m_p{R_i}\vec{v}_i,
\end{equation}
where the velocities are also measured with respect to the centre of
mass velocity. Finally the peculiar velocity of a halo is computed as
the centre of mass velocity.

The halo and environment used in this paper are publicly available through the
MultiDark data base\footnote{\texttt{http://www.multidark.org/MultiDark/}}. The
halo data is thoroughly described in \cite{2013AN....334..691R}.

\section{Web Finding Algorithms}
\label{sec:algorithms}

We use two algorithms to define the cosmic web in cosmological N-body
simulations. Both are based on the same algorithmic principle, which
determines locally a symmetric tensor which can be diagonalized to yield
three real eigenvalues $\lambda_1>\lambda_2>\lambda_3$ and their
corresponding eigenvectors $\hat{e}_{1}$, $\hat{e}_2$ and
$\hat{e}_3$. This allows for a local classification into one of the 
following four web types: void, sheet, filament and peak depending
on whether the number of eigenvalues larger than a given threshold
$\lambda_{th}$ is $3$, $2$, $1$ or $0$, respectively.

We use two different symmetric tensors. The first is the shear tensor,
defined as the Hessian of the gravitational potential, normalized in
such a way as to be dimensionless:

\begin{equation}
T_{\alpha\beta} = \frac{\partial^2\phi}{\partial
  r_{\alpha}\partial r_{\beta}},
\end{equation}
where $\phi$ is the gravitational potential rescaled by a factor $4\pi
G\bar{\rho}=3/2\Omega_m H_{0}^2$ in such a way that the Poisson
equation can be written as $\nabla^{2}\phi  = \delta$, where $\delta$
is the matter overdensity, $\bar{\rho}$ is the average matter density,
$H_{0}$ is the Hubble constant at the present time and $\Omega_m$ is the
matter density parameter. A detailed presentation of this algorithm
can be found in \cite{Tweb}.

The second tensor is the velocity shear:

\begin{equation}
\Sigma_{\alpha\beta} = -\frac{1}{2H_{0}}\left(\frac{\partial
  v_{\alpha}}{\partial r_{\beta}}+ \frac{\partial v_{\beta}}{\partial
  r_{\alpha}}\right),
\end{equation}
where the $v_{\alpha}$ correspond to the components of the peculiar
comoving velocities. With this definition the trace of the shear
tensor is minus the divergence of the velocity field normalized by the
Hubble constant $-\nabla\cdot {\bf}v /H_{0}$. A detailed description
of this algorithm can be found in \cite{Vweb}.

\subsection{Numerical considerations}

In this paper we compute the cosmic web on cubic grids of two different
resolutions $256^3$ and $512^3$ that roughly correspond
to scales of $1$ and $0.5\hMpc$, respectively. For the T-Web we
interpolate first the matter density field using a Cloud-In-Cell (CIC)
scheme. Then we smooth using a Gaussian kernel with a spatial variance
equal to the size of one grid cell. This smoothed matter density field is
transformed into Fourier space to solve the Poisson equation and find
the gravitational potential $\phi$. The Hessian is computed using a finite
differences method. Finally, the eigenvalues and eigenvectors are
computed on each grid point.

For the V-Web we interpolate first the momentum density field over a
grid using the CIC scheme and then apply a Gaussian smoothing with a
spatial variance of one grid cell. We use the matter density field,
which is also CIC interpolated and Gaussian smoothed, to normalize the
momentum field. This ratio between the momentum and matter density
field is what we consider as the velocity field to compute the shear
tensor on each grid point. In this case we also compute the
eigenvalues and eigenvectors on each grid point.

We caution the reader that the results reported by
\cite{Vweb} and \cite{Libeskind2013} use a velocity field that is
calculated by a Gaussian smoothing of the CIC velocity field without
taking into account any weight by mass.

\section{Our Numerical Experiments}
\label{sec:experiments}

In this paper we use the data and the methods described above to
perform two kinds of measurements: the preferential alignment (PA) and the
average value of the angle along the eigenvectors of interest.

We note that we measure alignments along the eigenvectors of the
cosmic web without defining first whether each point corresponds to a
filament or a wall. However, for simplicity and readability we describe
our results in terms of alignment with respect to filaments and
sheets. Given that the direction along a filament should be defined by the
eigenvector $\hat{e}_3$ corresponding to the smallest eigenvalue $\lambda_3$, a
strong alignment along that vector will be reported as an alignment
along filaments. Correspondingly, the first eigenvector $\hat{e}_{1}$
defines the direction perpendicular to walls, a strong alignment along
this vector will be reported as an anti-alignment along
walls. Finally, a strong alignment along the second eigenvector
$\hat{e}_2$ in company with an anti-alignment with $\hat{e}_1$ is
reported as an alignment with walls and anti-alignment with respect to
filaments.

We avoid the classification into filaments and walls for the following
reason. Partitioning the simulation into web elements implies
a choice regarding the value for the parameter $\lambda_{\rm th}$ in
two different web finders. This has been done before for each web
finder independently. However, we consider that deriving results
independent of the choice of parameters $\lambda_{\rm th,TWEB}$,
$\lambda_{\rm th, VWEB}$ provides clear data to understand the
connection of DM haloes with the cosmic web.  

A possible disadvantage is that we are mixing the alignment signal
of different environments. For instance, if half of the halo population of
fixed mass is aligned along filaments (the vector $\hat{e}_3$), and the
other half along planes (perpendicular to $\hat{e}_1$ but without a
clear signal along $\hat{e}_3$, $\hat{e}_2$), the total signal of the
alignment along $e_3$ might appear diluted in comparison to a signal
taken separately for filaments and walls. However, as it has been
shown in \cite{Libeskind2013} the alignment signals are robust across
different environments.

\subsection{Preferential Alignment}

The first measurement is a rough approximation to find out along which
axes haloes are aligned. We refer to this as PA.

We use the fact that for a given vector under study $\hat{s}$ and the
three eigenvectors the following identity holds 
 
\begin{equation}
(\hat{s}\cdot\hat{e}_1)^2 +(\hat{s}\cdot\hat{e}_2)^2 +(\hat{s}\cdot\hat{e}_3)^2 =1.
\end{equation}
Using this we know that all haloes can be split into three groups:

\begin{enumerate}
\item Haloes with $(\hat{s}\cdot\hat{e}_1)^2> (\hat{s}\cdot\hat{e}_2)^2$
  and $(\hat{s}\cdot\hat{e}_1)^2> (\hat{s}\cdot\hat{e}_3)^2$, which can
  be considered to aligned mostly along $\hat{e}_1$.
\item Haloes with $(\hat{s}\cdot\hat{e}_2)^2> (\hat{s}\cdot\hat{e}_1)^2$
  and $(\hat{s}\cdot\hat{e}_2)^2> (\hat{s}\cdot\hat{e}_3)^2$, which can
  be considered to aligned mostly along $\hat{e}_2$.
\item Haloes with $(\hat{s}\cdot\hat{e}_3)^2> (\hat{s}\cdot\hat{e}_1)^2$
  and $(\hat{s}\cdot\hat{e}_3)^2> (\hat{s}\cdot\hat{e}_2)^2$, which can
  be considered to aligned mostly along $\hat{e}_3$.
\end{enumerate}

If the halo population does not show any PA, then
all the haloes must be evenly distributed along these three
populations. On the contrary, if there is more than one third of the
halo population in one of these sets, then this will indicate a
PA along one of the axes. However, this statistics
does not give a precise answer on the degree of the alignment

\subsection{Average Alignment Angle}

We emphasize that we focus on the alignments with respect to the
eigenvectors regardless of the web type. We recall that the
eigenvector $\hat{e}_1$ is perpendicular to the plane defining a sheet
and the line describing a filament; and $\hat{e}_3$ is the vector that
marks the direction of a filament and lies on the plane of a
sheet. Therefore we focus on quantifying the degree of alignment along
these two eigenvectors. 

This experiment complements the results obtained by the PA statistic by
computing the average and standard deviation of
$\vert\langle\hat{s}\cdot\hat{e}_1\rangle\vert$ and
$\vert\langle\hat{s}\cdot\hat{e}_3\rangle\vert$.  We perform these
tests in different populations split into different mass bins
logarithmically spaced between $1\times 10^{9}$ and
$1\times10^{14}$\hMsun. 

In a separate test we make the same measurements but this time
splitting the halo sample by other properties such as:
circularity, concentration, local matter density, spin and
triaxiality. In this case we take the upper and lower $30\%$ of the
haloes according to each property and measure the strength of the
alignment by the average value of
$\vert\langle\hat{s}\cdot\hat{e}_1\rangle\vert$ and
$\vert\langle\hat{s}\cdot\hat{e}_3\rangle\vert$.

\section{Results}
\label{sec:results}

\subsection{Preferential Alignment}

\begin{figure*}
\includegraphics[width=0.8\textwidth]{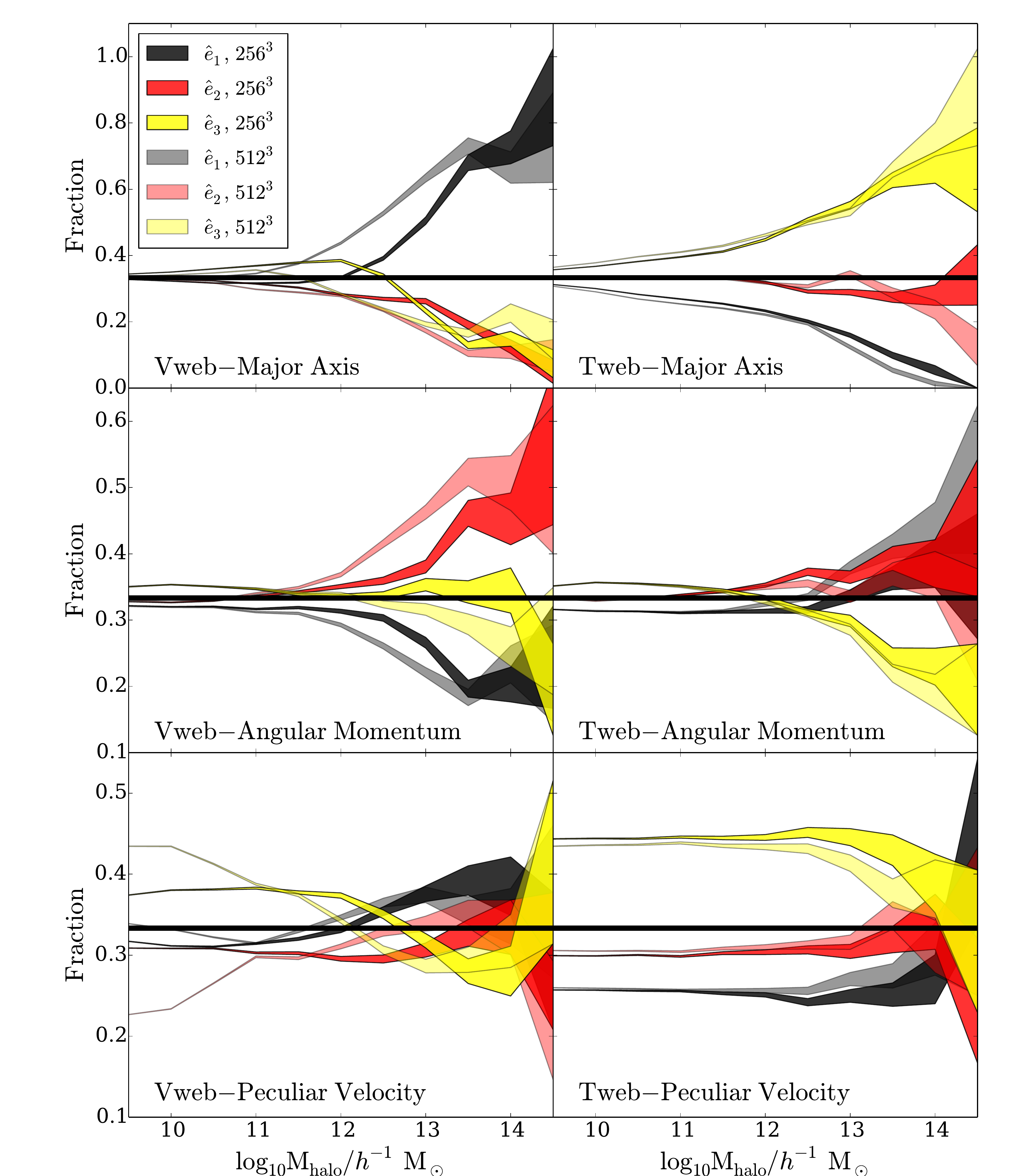}
\caption{Fraction of haloes in a mass bin that show a PA with respect to an eigenvector in the cosmic
  web; $\hat{e}_1$ (black) defines the direction perpendicular to a wall and
  $\hat{e}_3$ (yellow) indicates the direction along a filament. Each row
  presents one of the three properties studied in this paper: shape (major axis),
  angular momentum and  peculiar velocity. The left-hand (right-hand) column presents the
  results against the V-web (T-web). Strong colours refer to $256^3$ grid
  resolutions and lighter colours to a $512^3$ grid. The thick black
  horizontal line at $0.33$ corresponds to the expected fraction for a
  random vector field. The uncertainty is computed assuming Poissonian
  statistics in each mass bin.
\label{fig:preferential}}
\end{figure*}

Fig. \ref{fig:preferential} presents all the results for the
PA summarizing to a good extent the main
results of this paper.

The shape alignment and the V-web (upper row, left-hand column) give a
different perspective. First, there seems to be little evidence
for an alignment for masses below $10^{11}-10^{12}\hMsun$, depending
on the grid resolution. Secondly, the alignment at higher masses goes
along the first eigenvector $\hat{e}_{1}$ meaning that they mostly lie
perpendicular to the filaments and sheets. In the discussion section
we clarify this result that at first sight might seem puzzling.

For shape alignment and the T-web (upper row, right-hand column) we
find a strong PA along the third eigenvector $\hat{e}_{3}$. This  
signal increases steadily with mass and is almost independent of the
grid resolution. At high masses between $70$ and $100$ per cent of the
haloes have their major axis aligned along $\hat{e}_{3}$ which means
that they mostly lie along filaments and sheets.

The angular momentum in the V-web (middle row, left-hand column) presents 
a signal of alignment along the second eigenvector $\hat{e}_{2}$;
between $45$ to $60$ percent of the haloes are aligned along that
direction, while there is a minority of haloes aligned with
$\hat{e}_{1}$.  There is a clear change in trends around
$10^{11}-10^{12}\hMsun$ depending on the grid resolution;
below that mass range there is no evidence for alignment while at
higher masses all the trends we describe are noticeable. This means
that the angular momentum of haloes above $10^{12}\hMsun$ tends to lie
along walls, parallel to the vector $\hat{e}_2$; without any clear
trend for alignment/anti-alignment with respec to filaments.

For the angular momentum alignment and the T-web (middle row, right-hand
column) we find no evidence for any alignment at low masses $<10^{12}\hMsun$.
At higher masses $>10^{12}\hMsun$ there is a weak signal
of PA along the first and second eigenvectors;
between $35$ to $45$ per cent of the haloes are aligned with respect to
$\hat{e}_1$ and $\hat{e}_2$. Correspondingly, between $10\%$ to $20\%$
of the haloes are aligned along $\hat{e}_{3}$. This means
that most of the haloes are perpendicular to the filaments and do not
have a clear alignment with respect to walls.

The peculiar velocities (lower panels) show a weak but consistent
alignment along the third eigenvector $\hat{e}_{3}$ of the T-web for
all masses below $10^{13.0}\hMsun-10^{13.5}\hMsun$ depending on the grid
resolution. $45\%$ of the haloes are aligned this way, while only $25\%$ are
aligned along the first eigenvector $\hat{e}_1$. This suggests that
haloes tend to move along filaments and parallel to the walls, except
at higher masses where the alignments get randomized.  In contrast,
the peculiar velocities with respect to the V-web show the same,
although weaker, trend and only for low mass haloes $<10^{12}$.

In the next subsections we present a complementary account of these
results by showing quantitative results of the average angle
between vector pairs describing the alignments discussed so far.

\subsection{Shape Alignment}

\begin{figure*}
\includegraphics[width=0.75\textwidth]{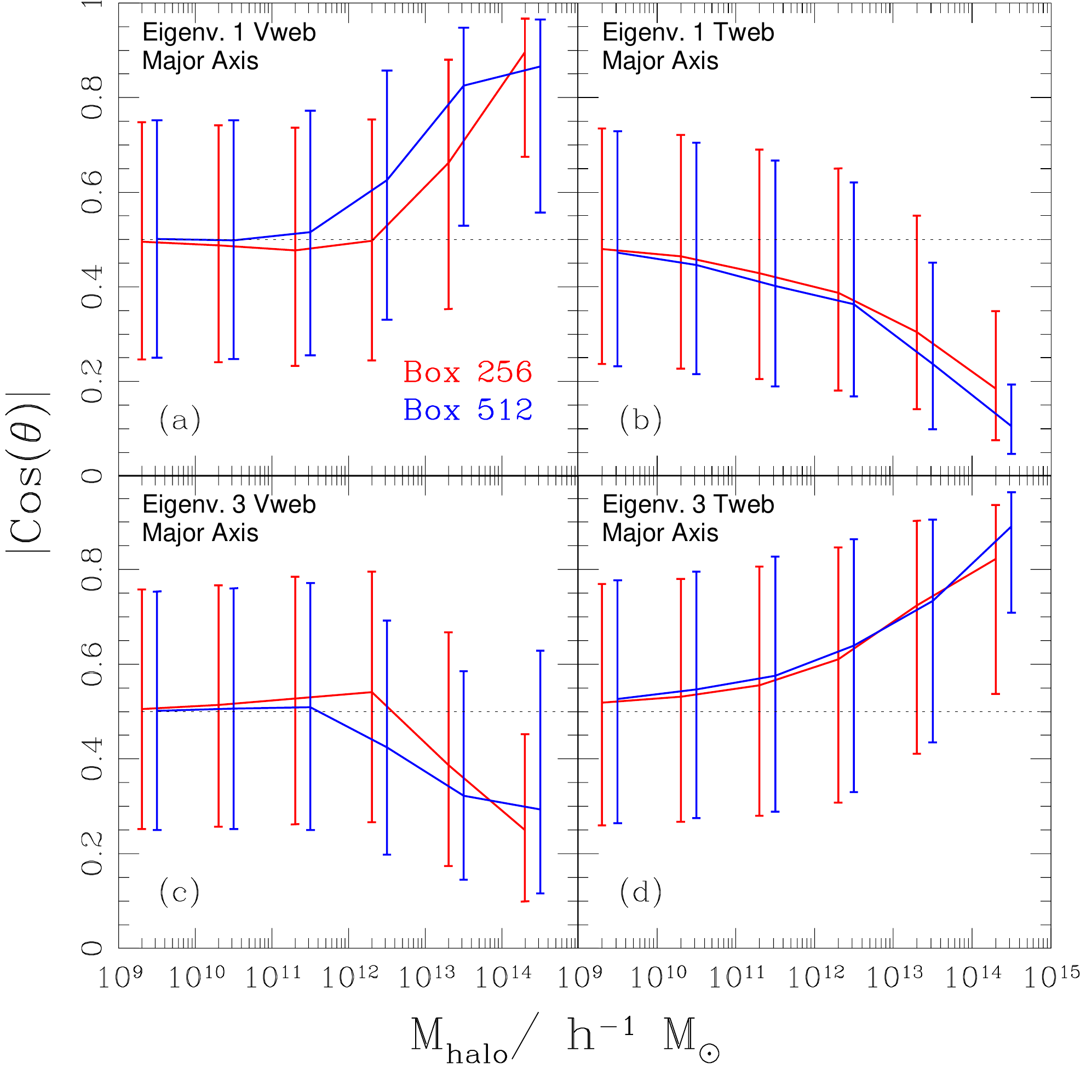}
\caption{Median of $|\cos\theta|$ quantifying the shape alignment for
  the V-web (left) and the T-web (right) at two different grid
  resolutions as a function of halo mass.  The errorbars show the $20$ and $80$ percentiles 
  of the distribution; the
  errors on the median are too small to be noticed in this figure. In the upper (lower) panels
  the angle $\theta$ is measured between the halo major axis and the first (third)
  eigenvector.\label{fig:shape_alignment} }
\end{figure*}

Fig. \ref{fig:shape_alignment} presents the main results for the
angles between the first and third eigenvectors and the major shape
axis as a function of halo mass.  Notice that the errorbars shown correspond
to the $20$ and $80$ percentiles; the actual error of the medians are
always small, $\sim 1-5$ percent, and therefore are not shown.  

In the case of the V-web (left-hand column) we have a clear alignment with
respect to the first eigenvector at high masses $>10^{12}\hMsun$, with
values $\muavg\approx 0.8$ well above the expected value of $0.5$ for a
random distribution. With respect to the third eigenvector we measure
an anti-alignment with $\muavg\approx0.3$. For low masses
$<10^{12}\hMsun$ we do not detect any alignment signal. This is
consistent with the PA results of massive haloes
perpendicular to filaments and parallel to walls.

The T-web (right-hand column) shows alignment trends starting at masses of
$10^{10}$\hMsun, two orders of magnitude below than the V-web. In this
case we measure an alignment along the third eigenvector and an
anti-alignment along the first eigenvector. In the latter case at the
highest masses $\muavg\approx 0.8$, while in the former at $\muavg\approx
0.2$. This strong alignment/anti-alignment signal mirrors the
interpretation from the PA results that describe haloes lying parallel
both to filaments and walls.

\subsection{Angular Momentum Alignment}

\begin{figure*}
\includegraphics[width=0.75\textwidth]{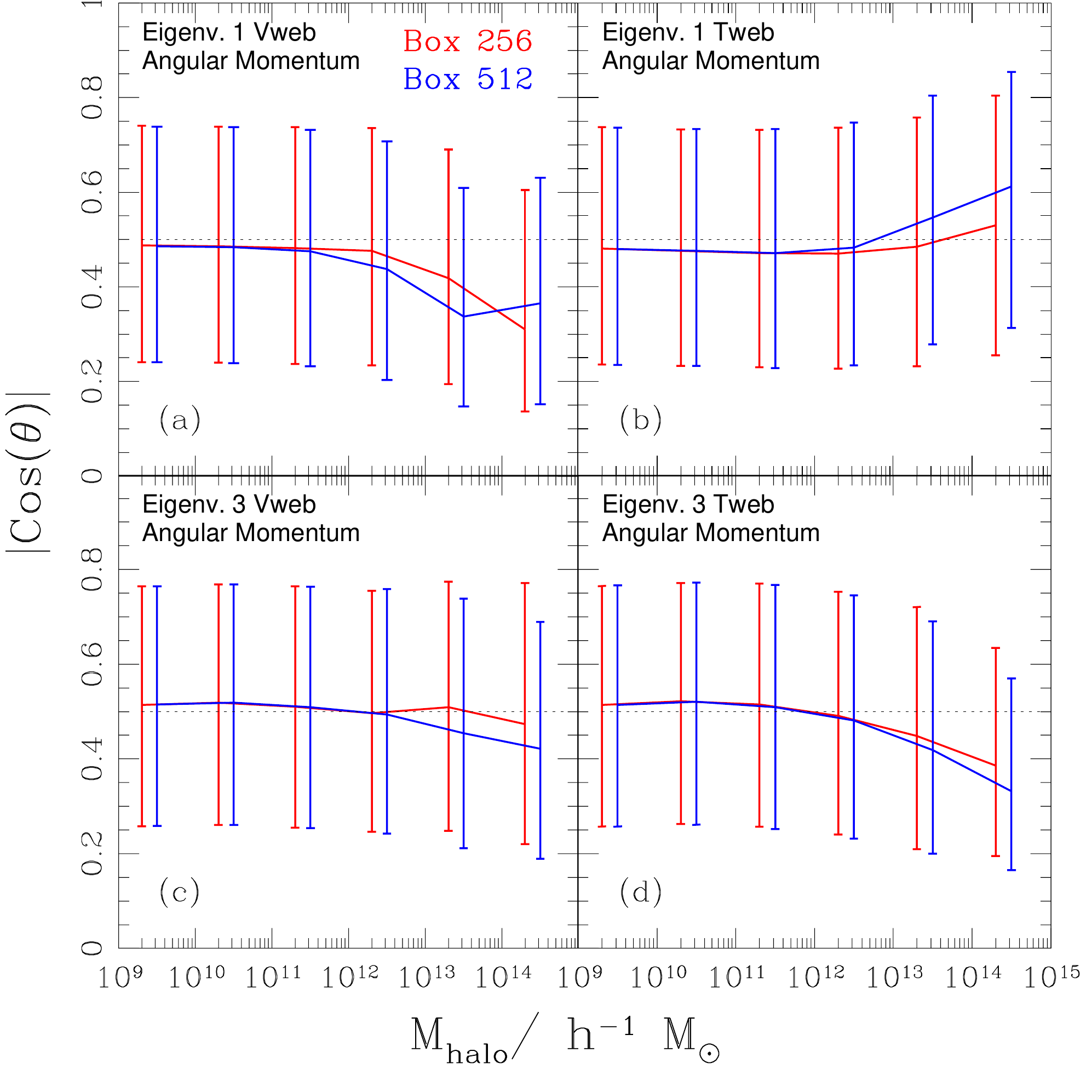}
\caption{Median of $|\cos\theta|$ quantifying the angular momentum
  alignment for the V-web (left) and the T-web (right) for two different
  grid resolutions as a function of halo mass. In the upper (lower)
  panels the angle $\theta$ is measured between the first (third)
  eigenvector and the angular momentum
  vector.\label{fig:spin_alignment}} 
\end{figure*}

We now focus our attention on the angular momentum alignment. Fig.
\ref{fig:spin_alignment} shows the results as a function of halo mass
following the same panel layout as in Fig.
\ref{fig:shape_alignment}. In all cases we see that these alignment
trends are weaker than the shape alignments. For the V-web low mass haloes
$<10^{12}$\hMsun do not show any PA with the
cosmic web. Haloes more massive than this threshold have their angular
momentum slightly perpendicular to the direction defined by the first
eigenvector and are uncorrelated with the third eigenvector. This translates
into a weak tendency for the angular momentum to lie parallel to walls.

In the case of the T-web, the alignment for low mass haloes $<10^{12}$\hMsun
is also absent. More massive haloes present a weak alignment along
first eigenvector and anti-alignment with the third eigenvector. This
provides a quantitative expression of the results derived from the
PA whereby the angular momentum is weakly
perpendicular to filaments.

\subsection{Peculiar Velocity Alignment}

\begin{figure*}
\includegraphics[width=0.75\textwidth]{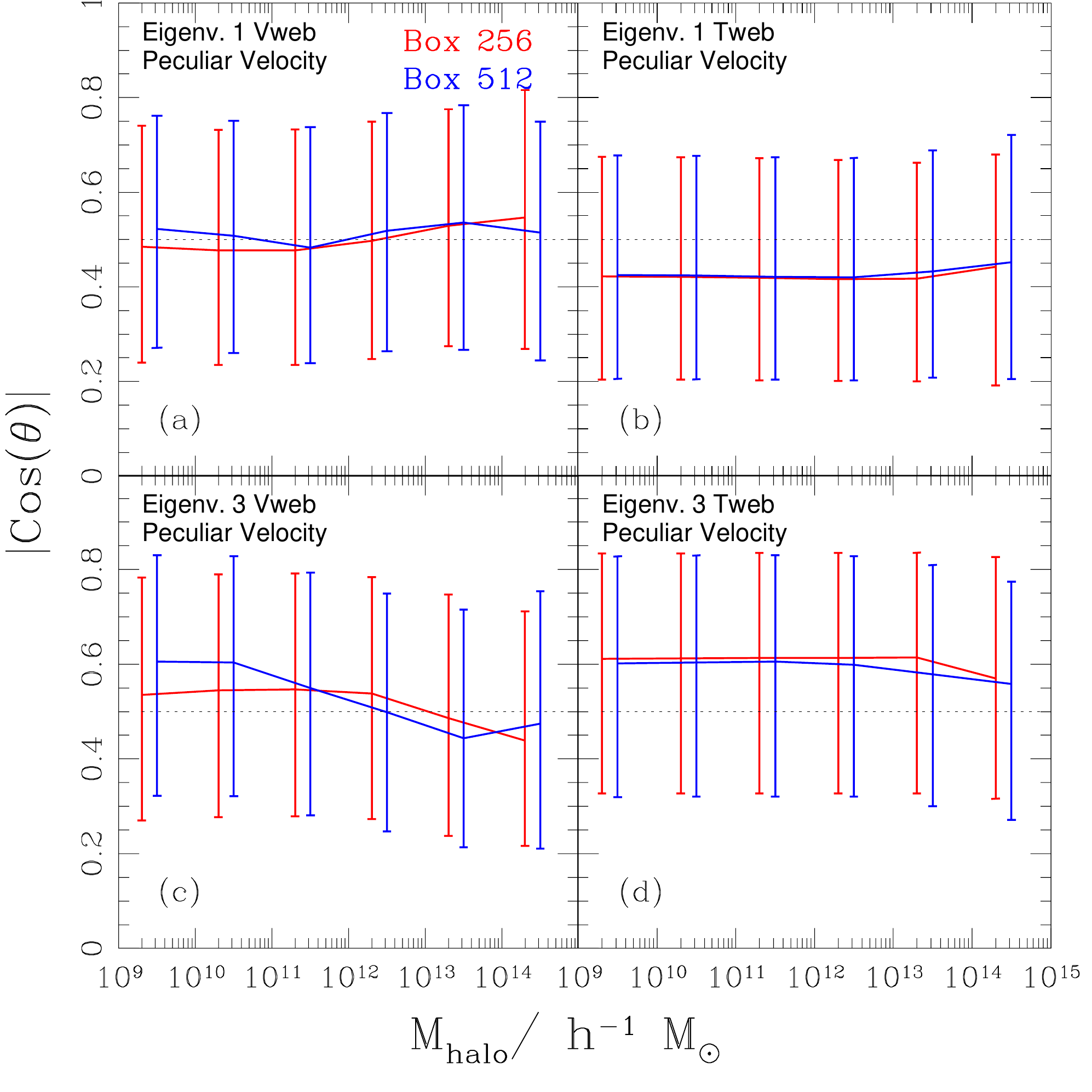}
\caption{Median of $\vert\cos\theta\vert$ quantifying the peculiar velocity
  alignment with the V-web (left) and the T-web (right) for two different
  grid resolutions as a function of halo mass. In the upper (lower)
  panels the angle $\theta$ is measured between the first (third)
  eigenvector and a halo's peculiar velocity.
\label{fig:velocity_alignment}}
\end{figure*}

Fig. \ref{fig:velocity_alignment} shows the results for peculiar
velocities alignments. In the case of the V-web, the peculiar
velocities show a weak signal of alignment ($\muavg\approx0.55$)
along the third eigenvector for low masses $<10^{12}$\Msun and a weak
anti-alignment at higher masses. The strength of the alignment also shows
a clear dependency on the grid size used to compute the web.

The T-web shows a stronger alignment with the third eigenvector at all
masses with $\muavg\approx0.6$ and an anti-alignment with the first
eigenvector with $\muavg\approx 0.4$. In contrast to the V-web results,
these trends remain basically unchanged at all masses and grid
resolutions, with only minor changes for haloes masses $>10^{13}$\hMsun.

\subsection{What drives the alignment}

We wish to understand what other selection criteria on halo
properties can produce a stronger local alignment for the shape, spin
and peculiar velocities. We split the halo population into low and
high mass haloes imposing a cut at $M_{\rm halo}=10^{11}$\hMsun. This
allows us to have robust statistics on the high mass end. We have also
computed these results for a cut at $M_{\rm halo}=10^{12}$\hMsun and
checked that the results we report below are not affected by this
change.

For each mass interval we perform cuts in the following properties:
halo spin, concentration, halo triaxiality defined as $(a^2-b^2)/(a^2-c^2)$ with
high (low) triaxiality corresponding to prolate (oblate) shapes,
circularity ($c/a$)  and halo inner density (virial mass divided by
volume out to the virial radius). We measure the web alignments in two
sets, each one including the $30\%$ of haloes in the lower/higher end
of the corresponding property. 

\begin{figure*}
\includegraphics[width=0.90\textwidth,angle=90]{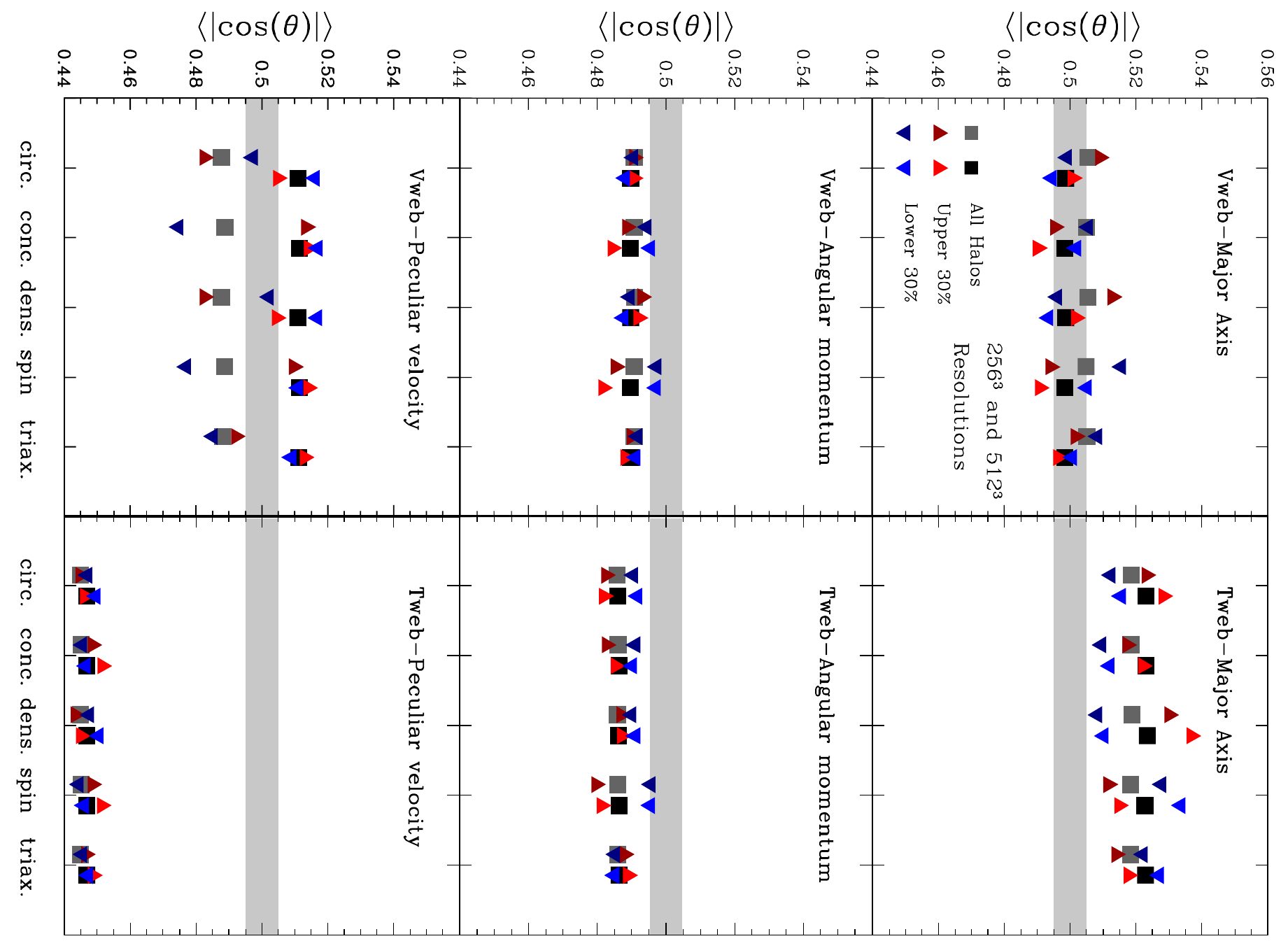}
\caption{Median of the $|\cos\theta|$ for the shape alignment of the
  major axis with the third eigenvector. Each panel shows
  different subsamples according to selections on five different
  properties: circularity, concentration, halo inner density, spin and
  triaxiality. Each subsets includes $30\%$ of haloes in the lower/higher end of each
  property. The grey band indicates the region around
  $|\cos\theta|=0.5$ that indicates the absence of alignment.}
\label{fig:drive}
\end{figure*}

Fig. \ref{fig:drive} shows the results for the major axes of
the V-web and T-web (left and right panels) and the shape major axis,
the angular momentum vector, and the halo peculiar velocity (top, middle
and bottom panels, respectively).  As here we show the average
of the halo population above the lower mass limit imposed, the halo
masses that dominate the statistics are close to this lower limit $~10^{11}\hMsun$.
We show the result for the T-web and V-web calculated using the two
available resolutions, but as can be seen in general we find no
significant differences in our results.

Haloes with higher circularity and inner density show
a higher alignment with the V-web major axes.  The opposite is the case
of haloes with higher concentration, spin and triaxiality.  This trend
is also visible in the angular momentum versus V-web major axis alignment only
for the concentration and spin, with little differences evidenced by the
other halo properties.  The alignment strengthening is somewhat
reversed when comparing the V-web with the halo peculiar velocity,
with some evidence for a strengthening with lower circularity and
inner density, and a weakening with spin and triaxiality.

On the right-hand panels the trends can also be readily seen.  The T-web
versus major axis alignments are stronger for higher circularity,
concentration and inner density, and weaker for higher spins and
triaxialities.  With respect to the angular momentum, the alignment
is weaker for higher circularities, concentrations, inner densities and
spins, and is only strengthened when the triaxiality is higher.
Not much difference is seen in the T-web versus halo peculiar velocity
alignments, being this the only regime where there is a clear difference
in influence of halo properties on the alignments with the T-web and V-web;
the latter do show important changes on the lower left-hand panel.

As can be seen, higher concentrations, circularities and densities,
as well as lower spins and triaxialities, produce a similar effect on the
T-web versus major axis alignments to that of diminishing the mass
of the haloes.  This has also been detected in other halo properties such
as their clustering amplitude in what has been termed assembly bias
\citep[][e.g.]{Gao2005,Li2008,Lacerna2011} where haloes of different
ages and equal mass show different clustering amplitudes, which could
also be interpreted as a change in the effective halo mass according
to halo properties.

An older halo age has been shown to come along with more spherical halo
shapes, higher concentrations and halo inner density.  However, in the
assembly bias scenario of \cite{Gao2005}, these haloes tend to show a clustering
amplitude consistent with that of the median halo population of higher masses.
In terms of the alignments with the T-web, our results show the opposite
trend.  This shows that the physics behind the clustering amplitude and
alignments do not necessarily coincide.  The full understanding of
the dependence of halo properties and/or on their environment is
a complex problem, but these results provide extra information that complements
that obtained from clustering measurements.

\subsection{Interweb Alignment}

\begin{figure}
\includegraphics[width=0.40\textwidth]{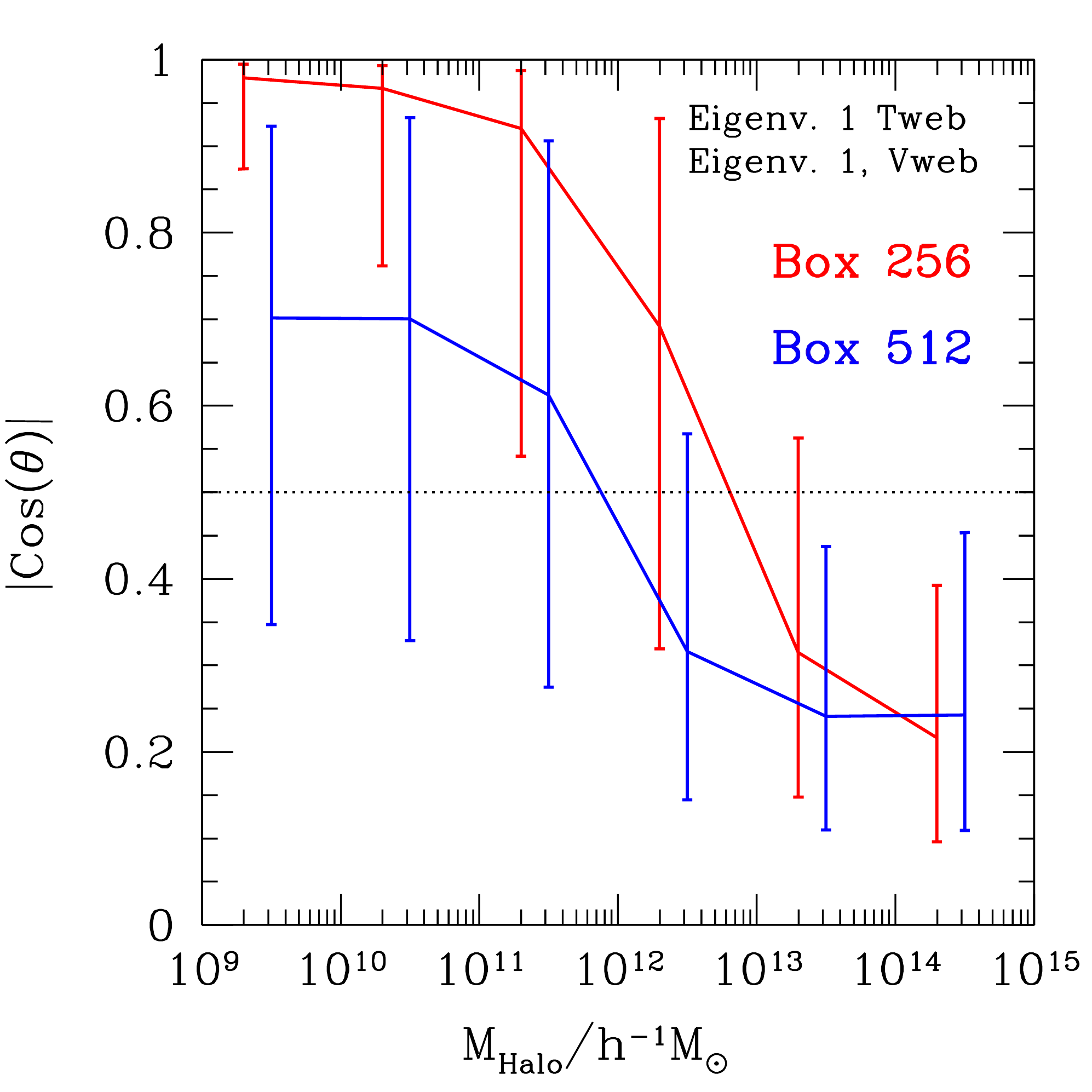}
\caption{Median of the interweb alignment for the two grid
  resolutions as a function of the DM mass corresponding to
  the haloes where the measurement was made. The error bars indicate
  the lower and upper quartiles.
\label{fig:interweb}}
\end{figure}

Perhaps the most striking result so far is that the two web algorithms give
different results for the alignment of massive haloes. This is not
completely unexpected given that the two algorithms are based on
different physical premises to obtain the directions defining the
eigenvectors. However, we investigate the origin of the different
alignment statistics by studying the inter-web alignment.
 
For the two algorithms, T-web and V-web, we have the information for their
eigenvectors and eigenvalues on exactly the same positions defined
by the grids. This allows us to compute the pair-wise alignment
between the eigenvectors in the two web finders.

We restrict our analysis to the grid cells that are occupied by
haloes. Otherwise, if we decided to perform this kind of analysis on
all the grid cells, the statistics would be more representative of the void
regions as they dominate in number the fraction of cells in the
simulation.

Fig. \ref{fig:interweb} shows the values for $\muavg$ between the
two $\hat{e}_1$ eigenvectors in the T-web and the V-web.  The Figure
shows that there is an alignment, $\muavg\approx 1.0$, for low mass
haloes and an anti-alignment, $\muavg\approx 0.2$ for massive ones.

The transitional scale is located around $(10^{11.5}-10^{12.5})$\hMsun
depending on the grid resolution. The coarse grid $(256^3)$ shows the
transition at higher masses than the fine grid $(512^3)$.   We also
note that the alignment is weaker in the finer grid, ($\muavg\approx
0.7$) than in the coarser grid ($\muavg\approx1.0$). 

These two facts (alignment at low masses and low grid resolution)
points towards an explanation in terms of the  linear / non-linear
growth of structure. When the alignment is present on linear scales
the divergence of the velocity field is proportional to the
overdensity, i.e. the trace of the shear field is proportional to the
trace of the tidal field.

On the scale where the haloes more massive than $10^{13}$\hMsun are
located, the relationship between the velocity shear and the tidal
field changes. There, the fastest momentum-weighted collapse direction
(defined by the V-web) is perpendicular to the direction where the
tidal compression is the highest. 

It is possible to consider that numerical effects can also
affect the estimation of the eigenvectors. For instance,
different levels of shot noise in the interpolation scheme to
construct the density and momentum grids could contribute to the
anti-alignment of the two algorithms. However, a detailed study of the
inter-web alignments is beyond the scope of this paper.

\section{Conclusions}
\label{sec:conclusions}

We have examined the alignment of shape, angular momentum and peculiar
velocity of DM haloes with respect to the cosmic web. We use
publicly available data from two algorithms implemented on a large
cosmological N-body simulation to study halo populations spanning five
orders of magnitude in mass. The first algorithm uses the tidal field (T-web)
and the second the velocity shear (V-web); both include results on
spatial scales of $0.5$ and $1.0$\hMpc.   

We quantify the alignments in two complementary ways. The first one
measures the fraction of haloes in a population that is preferentially
aligned with either one of the eigenvectors $\hat{e}_1$, $\hat{e}_2$
or $\hat{e}_3$ in the local definition of the cosmic web. The second
method measures the average value of the angle between an eigenvector
and the vector of interest. These two measurements give us a
complete picture for the different degrees of alignment in the web.

We find that the strongest alignment occurs for the halo shape with
respect to the T-web. In this case the haloes tend to align with the
third eigenvector, $\hat{e}_3$, meaning that they lie along filaments
and walls. This trend gets stronger as the halo mass
increases and agrees with all the results published so far. Instead,
for the momentum based V-web, there is only an anti-alignment for haloes
more massive than $10^{12}\hMsun$, a result that is presented here for
the first time. 

A much weaker alignment signal is present for the angular momentum. In
the T-web only the most massive haloes $>10^{12}$\hMsun are
perpendicular with respect to $\hat{e}_3$ (anti-aligned to filaments),
while for the V-web the massive haloes are aligned with $\hat{e}_2$,
lying along sheets. These results broadly agree with the
published literature. Nevertheless, in some publications
\citep{AragonCalvo2007,Hahn2007,AragonCalvo2014} there is an alignment signal
reported at lower halo masses $<10^{12}$\hMsun that we do not detect
in our measurements. Actually \cite{AragonCalvo2014}, using a
multiscale web-finding method, confirmed the mass $10^{12}$\hMsun as a
transitional scale between alignment /anti-alignment in filaments,
showing that it reflects the degree of evolution of the host
filament.

There are two possible explanations for the discrepancy. The first is
that in this low mass range the signal for the different
environments (mostly filaments and sheets) is mixed, diluting the
strength of the alignment. Another possible explanation can be
appreciated by carefully looking at the results for different
resolutions.  The $512^3$ grid does show a small anti-alignment for
low  mass haloes indicating that this signal could be related to scales
smaller than $1\hMpc$.  This would be consistent with
\cite{Paz2008} where, using a different technique, they find an
alignment for low mass haloes with the structure on small scales, and
an anti-alignment at large scales (the two-halo term). The diversity of
results could be then interpreted as a high sensitivity of the
alignment signal to the small scale cosmic web description, including
numerical choices as to how the relevant fields are interpolated.

A new result from our study is the alignment for the peculiar
velocities. Here we find a relatively strong signal of alignment along
the direction defined by the third eigenvector, $\hat{e}_3$, and
perpendicular to the first, $\hat{e}_1$. This signal is clear in the
T-web for all masses below $<10^{13}$\hMsun. This can be interpreted as
a flow parallel to walls and filaments. In the case of the V-web
similar signal, albeit weaker, is present only for the low mass haloes
$<10^{12}\hMsun$. A similar result was obtained by \citep{Padilla2005}
who found that peculiar velocities are larger in the direction
parallel to void walls. 

The different behaviour for the alignments of massive haloes in the
T-web and the V-web is explained by an anti-alignment between the
eigenvectors in the two web grids for massive haloes
$>10^{12}\hMsun$. For low mass haloes the directions defined by the two
webs point in the same direction. This trends can be interpreted as
non-linear effects that appear in the two different physical
descriptions for the cosmic web, but the impact of the numerical
choices to build the interpolation must be studied as well.

We also performed a simple study to find evidence of halo properties,
other than mass, in driving the alignments. We find that in the case of
shape, high concentration haloes or haloes with a low value of
the reduced spin parameter tend to show a stronger signal. This trend is
more pronounced in the T-web than in the V-web. Concerning angular
momentum we find that the anti-alignment signal is stronger for haloes
with high spin values. In the peculiar velocities we do not find any
effect in the T-web alignments, and the results for the V-web show 
wide variations with grid resolution that impedes driving any strong
conclusion.

Our study has shown that the alignment properties of DM
haloes depend on the physical definition of the cosmic web. This shows
that the main aspect of non-linear gravitational collapse might be
revealed easily depending on the choice of the physical context
(e.g. tidal field versus velocity shear). There is not a better method,
but different perspectives. In future work we will explore the
potential of these two complementary techniques in understanding the
environmental dependence of galaxy evolution.

\section*{Acknowledgements}

JEF-R acknowledges the financial support by Vicerrector\'ia de
Investigaciones at Universidad de los Andes through a FAPA grant.
NP acknowledges support by Fondecyt Regular No. 1110328 and BASAL PFB-06 "Centro de Astronomia y Tecnologias Afines".
SC acknowledges support from a PhD Fellowship from CONICYT
NP and SC acknowledge support by the European Commissions Framework
Programme 7, through the Marie Curie International Research Staff
Exchange Scheme LACEGAL (PIRSES-GA-2010-269264). NP thanks the
hospitality of the Max Planck Institute for Astrophysics at Garching
during his sabbatical year 2013-2014. The Geryon cluster at the Centro de Astro-Ingenieria UC was extensively used for the calculations performed in this paper.
The Anillo ACT-86, FONDEQUIP AIC-57, and QUIMAL 130008 provided
funding for several improvements to the Geryon cluster.

We thank Kristin Riebe for her work to make the cosmic web data
available in the Multidark data base. 

We thank the referee, Miguel Arag\'on-Calvo, for a constructive report
that helped us to improve the paper.

\section*{Appendix A. Detailed description of previous theoretical results}

In this Appendix we review the results that use a similar exploration
techniques for halo alignments with the cosmic web. Other kind of
alignment statistics based based on shape measurements
\citep{Basilakos2006} or modifications of the correlation
\citep[e.g.][]{Paz2008,Faltenbacher2009} that go beyond a local
computation and are not reviewed here. 

\begin{itemize}

\item
\cite{Libeskind2013}

They study the shape and angular momentum alignments with the cosmic
web defined by the velocity shear tensor method described in this
paper.  \cite{Libeskind2013} used the Bolshoi simulation and the halo
catalogs we use in this work. Results are reported for three mass bins $M_{\rm
 vir}<10^{11.5}$\hMsun, $11^{11.5}<M_{\rm vir}<12^{12.5}\hMsun$ and
$M_{\rm  vir}>12^{12.5}\hMsun$. The identification of the cosmic web
is done on a grid of $256^3$ with a Gaussian smoothing of $\sim
1$\hMpc over the velocity field. The way they compute this smoothed
velocity field differs from our computation. We do it based on the
momentum density field while \cite{Libesking2013} do not take into
account the mass in each cell.

The alignment signal for the angular momentum is weak while the shape
alignment signal is very strong. The shape alignment is such that the
eigenvector corresponding to the smallest eigenvalue is aligned with
the major axis. This effect is stronger for more massive haloes.  In
other words the major axis of a halo is aligned with a filament, and
lies on the plane that define a sheet. The angular momentum is
anti-aligned with the filament for massive haloes and weakly aligned
for low mass haloes. 

\item
\cite{Trowland2013}

They used the Millennium Run, which has $2160^3$ particles in a volume
of $500$\hMpc on a side. This corresponds to a particle mass of
$8.6\times 10^{8}$\hMsun. The catalog uses both haloes and subhaloes
identified with SUBFIND. Only haloes with more than 500 particles were
kept to get a robust computation for the spin. The angular momentum is
defined as the sum of the angular momentum of each particle with
respect to the centre of mass.

The method to define the filamentary structure is based on the
eigenvalues of the hessian of the density.  However, the analysis is
performed on a box of $300$\hMpc on a side. Four different Gaussian
smoothing scales are used: $2.0$, $3.0$ and $5.0$\hMpc.

By fitting the following functional form tothe $\cos(\theta)$ distribution

\begin{equation}
P(\cos\theta) =
(1-c)\sqrt{1+\frac{c}{2}}\left[1-c\left(1-\frac{3}{2}\cos^{2}\theta
  \right)\right]^{-3/2},
\label{eq:distro}
\end{equation}
they are able to quantify the degree of alignment ($c<0$) or
anti-alignment ($c>0$).  This parameterization is based on theoretical
expectations of Tidal Torque Theory (TTT) \citep{Lee2005}. At $z=0$,
the reported value is $c = −0.035 \pm 0.004$, where the uncertainty
was calculated using bootstrapping and resampling.

When the halo sample is divided between low mass and high mass haloes
with a transition scale $M_{\star}=5.9\times 10^{12}$\Msun, there is
a weak alignment signal of the angular momentum against the
principal filament axis for haloes above that mass, for haloes below
that scale there is a weak anti-alignment.

\item
\cite{Codis2012}

They study the alignment of the angular momentum dark relative to
the surrounding large scale structure and to the tidal tensor
eigenvalues.

They use a dark matter simulation with $4096^3$ DM particles in a
cubic periodic box of $2000$\hMpc on a side, which corresponds to a
particle mass of $7.7\times 10^9$\Msun. Haloes are identified
using a FoF algorithm with a linking length of 0.2 keeping all haloes
with more than 40 particles, which sets the minimum halo mass to be
$3\times 10^{11}$\Msun. In their work the particles were sampled on a
$2048^3$ grid and the density field was smoothed with a Gaussian
fileter over a scale of $5$\hMpc corresponding to a mass of $1.9\times
10^{14}$. The skeleton was computed over $6^{3}$ overlapping subcubes
and then reconnected.

The filament finder algorithm is based on Morse theory and defines a
Skeleton to be the set of critical lines joining the maxima of the
density field through saddle points following the gradient \citep{Sousbie2008b}. They also
compute the hessian of the potential over the smoothed density field
to get their eigenvectors.

The angular momentum of the halo is defined as $m_{p}\sum_{i}(r_i-\bar{r})\times
(v_i-\bar{v})$ where $\bar{r}$ is the centre of mass of the halo and
$\bar{v}$ is the average velocity.

They measure the alignment with each one of the eigenvectors. With
repecto to the minor eigenvector $\hat{e}_{3}$  (the filament direction) there is anti-alignement for masses $M>5\times10^{12}$\Msun and alignment for masses
$<5\times 10^{12}$\Msun; with respect to the intermediate eigenvector $\hat{e}_2$
there is a strong alignment at high masses and no alignment for low
masses; with respect to the major eigenvector $\hat{e}_{1}$
(normal to the wall plane) there is an anti-alignment signal at all
masses. The results from the Skeleton algorithm are in
agreement with the results from the Tidal web.  The transitional mass
is weakly dependent on the smoothing scale, varying between $1-5\times
10^{12}$\hMsun for smoothing scales between $1.0-5.0$\hMpc.

\item
\citep{Zhang2009}

They studied the angular momentum and shape alignment against
filaments.  They used a dark matter simulation with $1024^3$ DM particles in a
periodic box of $100$ \hMpc on a side. The particle mass is
$6.92\times10^{7}$\hMsun. Dark matter haloes are found using a FOF
algorithm with a linking length of 0.2 times the interparticle
distance. Only haloes with more than 500 particles are retained for
further analysis. The angular momentum is measured with positions
repect to the centre of mass and the shape is determined using the
non-normalized moment of inertia tensor.

The environment is found using the hessian of the density. The density
field was interpolated over a $1024^3$ grid and then smoothed with a
Gaussian filter of scale $R_{s} = 2.1$\hMpc. There are two methods to
define the direction of a filament. The first method uses the
eigenvalues of the hessian density; they take the filament
direction to be the eigenvector corresponding the single positive
eigenvalue of the hessian. The second method used a line that
connects the two terminal haloes in a filament segment.

For the method that uses the eigenvectors, they find that the strength
of the angular momentum alignment decreases with halo mass. For the
shape they study the alignment of the major axis with the
filament. The find an alignment signal in all mass bins, with an
stronger effect for more massive haloes. 

In a final experiment they measure the angular momentum alignment in four
different samples split by the strength of the shape alignment. They
find that haloes anti-aligned in shape, show a strong angular momentum
correlation; and a strong angular momentum anti-alignment for haloes
with a strong shape alignment.

\item
\citep{AragonCalvo2007}

They used the Multi-Scale Morphology Filter \citep{AragonCalvo2007} to describe th filamentary
structure. The method is based on the Hessian matrix of the density
field, which is computed from the particle distribution using a
Delaunay tesselation field estimatior (DTFE). This allows them to
identify clusters, filaments and walls.

They used a simulation with $512^3$ particles in a cubic box of $150$\hMpc. The
mass per particle is $2\times 10^{9}$\hMsun.  Halo identitification is
done with the HOP algorithm. They keep haloes with more than $50$
particles and less than $5000$, defining a mass range of$1-100\times
10^{11}$\hMsun. The principal axes of each halo are computed from the
non-normalized inertia tensor. The inertia tensor and the angular
momentum are computed with respect to the centre of mass of the halo. 

They compute two angles, one with respect to the direction
defining the filaments and the other the walls. Their results make a
distinction between haloes of more massive and less massive than
$10^{12}$\hMsun. The angular momentum tends to lie along the plane of
the wall, with a stronger alignment for massive haloes. The effect for
filaments is weaker, low mass haloes tend to align along the filament,
while high mass haloes tend to be anti-aligned. 

For the shape they find a very strong alignment along filaments. In
walls the major axis lies along the wall. Both alignments are stronger
for massive haloes.

\item

\citep{Hahn2007}

They used the hessian of the gravitational potential \footnote{In \citep{Hahn2007} the authors use a definition of the tidal field tensor equivalent with the T-Web method} applied on three simulations each of $512^3$
particles, with sizes $L_1=45$\hMpc, $L_2=90$\hMpc and
$L_3=180$\hMpc, this corresponds to particle masses of $4.7, 38.0,
300\times 10^7$\hMsun. Halo identification was done with a FOF
algorithm with 0.2 times the interparticle distance. They considered
haloes of at least 300  particles.

The web is obtained for a grid of $1024^3$ cells, the density field is
obtained with a CIC interpolation and smoothed using a Gaussian
Kernel. All the results correspond to a smoothing scale of
$R_{s}=2.1$\hMpc.

They report on the angle between the halo angular momentum vector and
the eigenvector corresponding to perpendicular directions to the
sheets and the direction of the filaments. This is divided into two halo
populations according to mass; low mass $5\times 10^{10} - 1.0\times 10^{12}$ and
high mass $>10^{12}$. They find a weak anti-alignment for filaments
and a stronger anti-alignment in the case of the sheets. For the
sheets the effect is stronger for the massive bin. The anti-alignment
along filaments is weak regardless of the mass. They do not report any
other significant statistic, but recognize that they suffer from
small-number statistics in voids.

\end{itemize}

\bibliographystyle{mn2e}

\end{document}